\documentclass[aps,prl,twocolumn,groupedaddress]{revtex4-2}
\usepackage{graphicx}

\begin{document}


\title{Optical resonances in silica-nanosphere dimers and trimers probed with high-energy electrons at the nanoscale}


\author{Martin Couillard}
\email[]{martin.couillard@nrc.ca}
\affiliation{National Research Council Canada, Energy, Mining and Environment Research Centre, 1200 Montreal Road, Ottawa, ON Canada K1A OR6}


\date{\today}

\begin{abstract}
Optical dielectric resonances have the ability to modulate optical energy, and by coupling multiple resonators into “photonic molecules”, new opportunities emerge in spectral engineering.  Using electron-beam spectroscopy, in a 300-keV scanning transmission electron microscope, we probe optical modes of photonic molecules at the nanometer scale.  Monochromated electron energy-loss spectra of 400nm-silica-sphere monomers, dimers and trimers display series of sharp peaks in the ultraviolet region.  For monomers, the peaks are assigned to low angular number optical whispering gallery modes.  For dimers and trimers, experimental observations, supported by simulations, reveal energy shifts and peak broadening resulting from coupling between spherical modes.  Variations in spectra for different electron probe positions demonstrate the possibility to extract information on local photonic density of states at the nanometer scale.
\end{abstract}


\maketitle

Advances in electron-beam spectroscopies performed in an electron microscope with a sub-nm probe have enabled unique access to nanophotonic properties \cite{polman2019,deAbajo2010}.  With electron energy-loss spectroscopy (EELS), plasmon modes have been mapped at the nanometer scale \cite{nelayah2007,rossouw2011}, even in three dimensions \cite{nicoletti2013}.  Radiative modes resulting from the Cherenkov effect are also detected in EELS \cite{kordahl2021,couillard2010,arslan2009,yurtsever2008} and cathodoluminescence \cite{polman2019}, often preventing bandgap measurements \cite{couillard2007,stoger2006, couillard2008} but allowing for the spatially-resolved analysis of photonic crystals \cite{cha2010, peng2019}.  Driven by recent interests in all-dielectric nanophotonics \cite{kuznetsov2016}, electrons-beam has also been applied to probe optical modes in dielectric (non-plasmonic) resonators using both EELS \cite{hyun2008} and cathodoluminescence \cite{muller2021,coenen2013}.  

Two distinctive type of optical resonances are predicted for an individual sphere: the plasmonic and the dielectric resonances \cite{tzarouchis2018}.  Nano-optics is usually associated with plasmonic resonances, but the second type of polarization also confines optical energy, and often with lower losses.  Whispering gallery modes (WGMs) are a subset of dielectric resonances \cite{oraevsky2002}.  From a geometrical optics perspective, such modes travel inside the sphere through total internal reflections, forming standing waves.  Electromagnetic fields for WGMs are highly localized near the surface of the sphere.  By bringing spheres in close proximity, coupling of such optical modes is therefore expected, and new structures can be created, such as photonic crystals or, for a limited number of resonators, photonic molecules \cite{boriskina2010}.

In this letter, we evidence coupling of whispering gallery modes in dimers and trimers composed of silica-quasispheres with diameters $\sim$400 nm by positioning a nanometer-scale electron probe at multiple locations and measuring energy losses of incident electrons.  We also demonstrate experimentally that different coupled WGMs are excited for different electron probe positions, pointing to an access to the local photonic density of states.   

EELS acquisitions have been carried out with a 300 keV FEI Titan cubed scanning transmission electron microscope (STEM) equipped with a monochromator and a Gatan Imaging Filter (Tridium 866ERS).  The microscope optics include an aberration corrector for the probe-forming lenses allowing for atomic resolution \cite{couillard2021} and atomic sensitivity \cite{couillard2011} capabilities.  The convergence and collection semiangles were set at 18 and 4.5 mrad, respectively.  With the monochromator, the EELS energy spread was less than 100 meV, while the spatial resolution was estimated at $\sim$1 nm.  To reduce the noise, 100 spectra were acquired over a region of 2nm x 2nm with an acquisition time of 30ms and a dispersion of 0.01eV/channel.  The spectra were then aligned in energy and averaged.  Samples were prepared by dispersing SiO$_2$ spheres (Nanocomposix NanoXact) in distilled water, and depositing a drop of the solution onto ultrathin ($<$3nm) carbon film supported on thicker lacey carbon film (Ted Pella).  Boundary element method (BEM) simulations were carried out with the Matlab package MNPBEM \cite{hohenester2012}.  This approach is based on a full relativistic formalism for arbitrary-shaped dielectrics, and incorporates EELS simulations \cite{hohenester2014}.  Dielectric constants used in the simulations were from Ref. \cite{palik1998}.  Simulations for a sphere are consistent with the modes predicted by an analytical solution \cite{deAbajo1999}, which was used in our previous study\cite{hyun2008}.  With BEM, the analysis could be extended beyond individual spheres to dimers and trimers.

 \begin{figure}
  \includegraphics[width=0.35\textwidth]{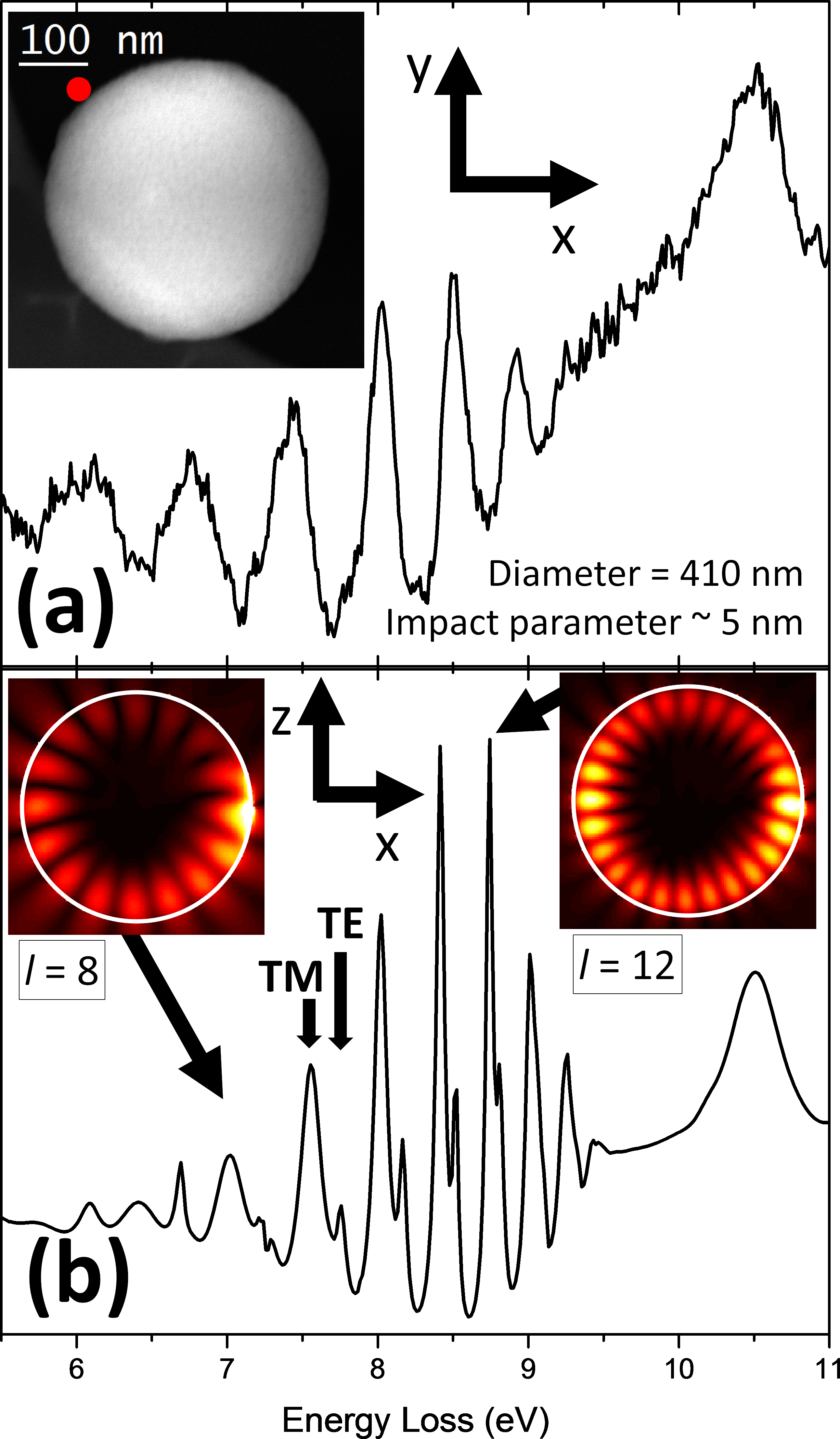}%
 \caption{\label{} (a) Experimental and (b) Simulated electron energy-loss spectra for an electron probe located $\sim$5nm beside a 410 nm silica sphere.  Inset in (a) displays a STEM image of the silica nanoparticle with the location of the electron probe (red dot).  Insets in (b) display $\vert H \vert$ maps for $l=8$ and $l=12$ on the \textit{x-z} plane (see text) with an electron travelling on the right of the sphere.}
 \end{figure}

Figure 1(a) displays an energy-loss spectrum acquired with an electron probe located $\sim$5nm from the surface of a 410 nm silica sphere, as shown in the inset STEM image.  Below the $\sim$9 eV SiO$_2$ bandgap, where no electron energy losses would be expected in the classical picture ($Im(-1/ \epsilon) \sim 0$), the spectrum instead displays a series of sharp peaks ascribed to optical whispering gallery modes \cite{hyun2008}.  The width of the peaks decreases as their energy position increases, until their position is close enough to the bandgap energy and their width start increasing again.  As presented in fig. 1(b), BEM simulations also display a series of peaks.  The simulated peaks are much sharper and, in contrast to our previous work \cite{hyun2008}, their position do not match with the experiments.  Multiple factors may explain these discrepancies, such as differences in morphologies or optical constants.  Due to limited information, we do not conclude on the exact nature of these discrepancies.

Optical modes in a sphere can be described by radial ($N$), angular ($l$), and azimuthal ($m$) numbers, along with a polarization $p$ (transverse electric (TE) or transverse magnetic (TM)).  WGMs corresponds to modes with  $N=1$ and $m=l \gg 1$.  The modes observed in EELS are predominantly TM in nature, with a smaller TE contribution, as labelled in Fig. 1(b) for one set of peaks.  For the transverse magnetic (TM) polarization, the energy of WGMs can be approximated as \cite{oraevsky2002}:

\begin{equation}
\omega_{l} \approx \frac{c}{a(\epsilon \mu)^{1/3}} \left[ \nu + 1.85576 \: \nu ^{1/3} - \frac{1}{\epsilon}\left( \frac{\epsilon \mu}{\epsilon \mu - 1} \right) ^{1/2} \right]
\end{equation}


where $\nu = l+1/2$ and $c$ is the speed of light.  Assuming a magnetic permeability $\mu \approx 1$, and knowing the sphere's diameter ($a$), the optical constants ($\epsilon$) as a function of energy, and the energy positions of peaks ($\hbar \omega_{l}$), it is then possible to assign $l$ to each simulated modes.  In the inset of fig. 1(b), two $\vert H \vert$ field distributions are displayed on an \textit{x-z} plane that includes both the center of the sphere and the electron trajectory.  The maps are for the induced fields.  We observe a ring-like pattern that is characteristic of WGMs, with 16 maxima for $l=8$ and 24 maxima for $l=12$.  As $l$ decreases, fields extend closer to the sphere's center and farther outside of the sphere.

 \begin{figure}
  \includegraphics[width=0.45\textwidth]{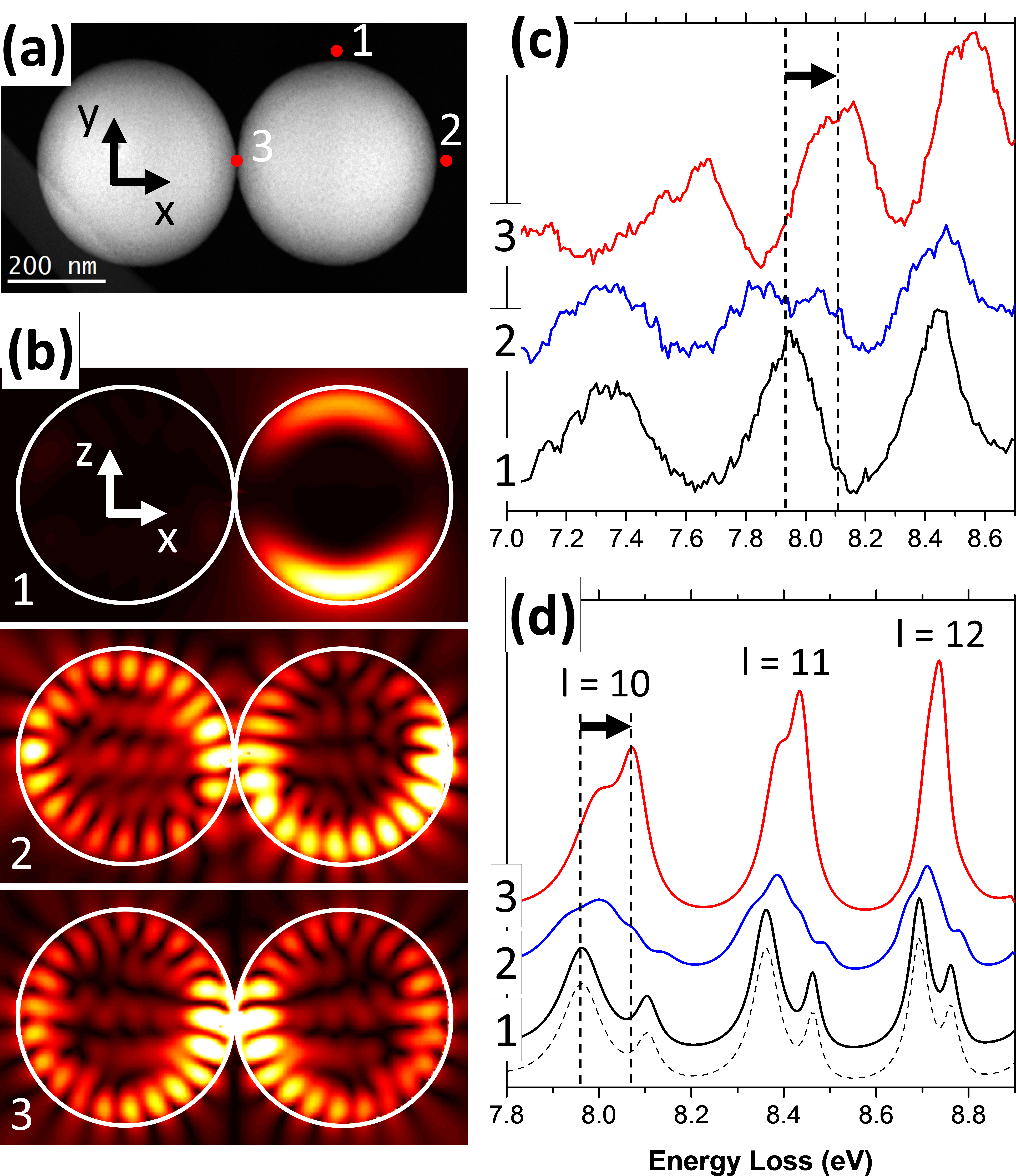}%
 \caption{\label{} (a) STEM image of silica bisphere with electron probe locations for $\vert H \vert$ maps displayed in (b), experimental spectra displayed in (c) and simulated spectra displayed in (d).  A simulated spectrum for an individual sphere is shown as a dotted line in (d).  $\vert H \vert$ maps are displayed on the \textit{x-z} plane that includes both center of the spheres.  Locations 1 and 2 are $\sim$5nm from the sphere's surface}
 \end{figure}

For a dimer (Fig.2(a)), spectra collected at three probe locations are displayed in (c), along with their corresponding simulated spectra in (d).  The spheres have similar diameters (left: 415 nm, right: 412 nm) and appear to be in contact, or possibly separated by only a thin surface coating.  Beam damage prevented the acquisition of spectrum images, so point analysis was used instead.  First, for the probe location 1 (see labels in (a)), the EELS spectrum closely resembles the spectrum for an individual particle.  In contrast, a spectrum acquired at location 2 displays peaks at similar energy, but significantly broadened.  Finally, for location 3, we observe blue-shifted peaks.  The shift in energy increases as the mode energy decreases.  Although the energy positions are not well reproduced, as explained above for monomers, BEM simulations recreate the trends experimentally observed for the dimer (Fig. 2(d)). Also, the magnitude of the shift in energy in the BEM simulations for position 3 is close to the one observed experimentally.  For instance, the energy shift for a peak at $\sim$8eV is $\sim$0.16 eV in the experiments and $\sim$0.12 eV in the simulations (see arrows in (c) and (d)).  In both cases, this shift increases for lower energy peaks.  Finally, the simulated spectrum for location 2 displays additional features that are not resolved experimentally, but are also indicative of optical coupling.  

Simulated $\vert H \vert$ fields distributions for $l=10$ are presented in Fig. 2(b).  The energy selected for each simulations corresponds to the peak maximum in each spectrum.  The distributions are shown on an \textit{x-z} plane that includes both centers of the spheres, and is parallel to the electron trajectory.  First, for location 1, fields are contained almost entirely within one sphere, producing a spectrum similar to the one for an individual sphere.  For location 2, the map displays fields that extends to both spheres.  Finally for location 3, for a beam at the center of the dimer, the resonance also produces fields in both spheres, but with a distribution that differs markedly, in particular around the spheres’ equator.  In addition, when compared to monomers, more fields are found closer to the spheres’ center and spill outside of the spheres.

 \begin{figure}
  \includegraphics[width=0.45\textwidth]{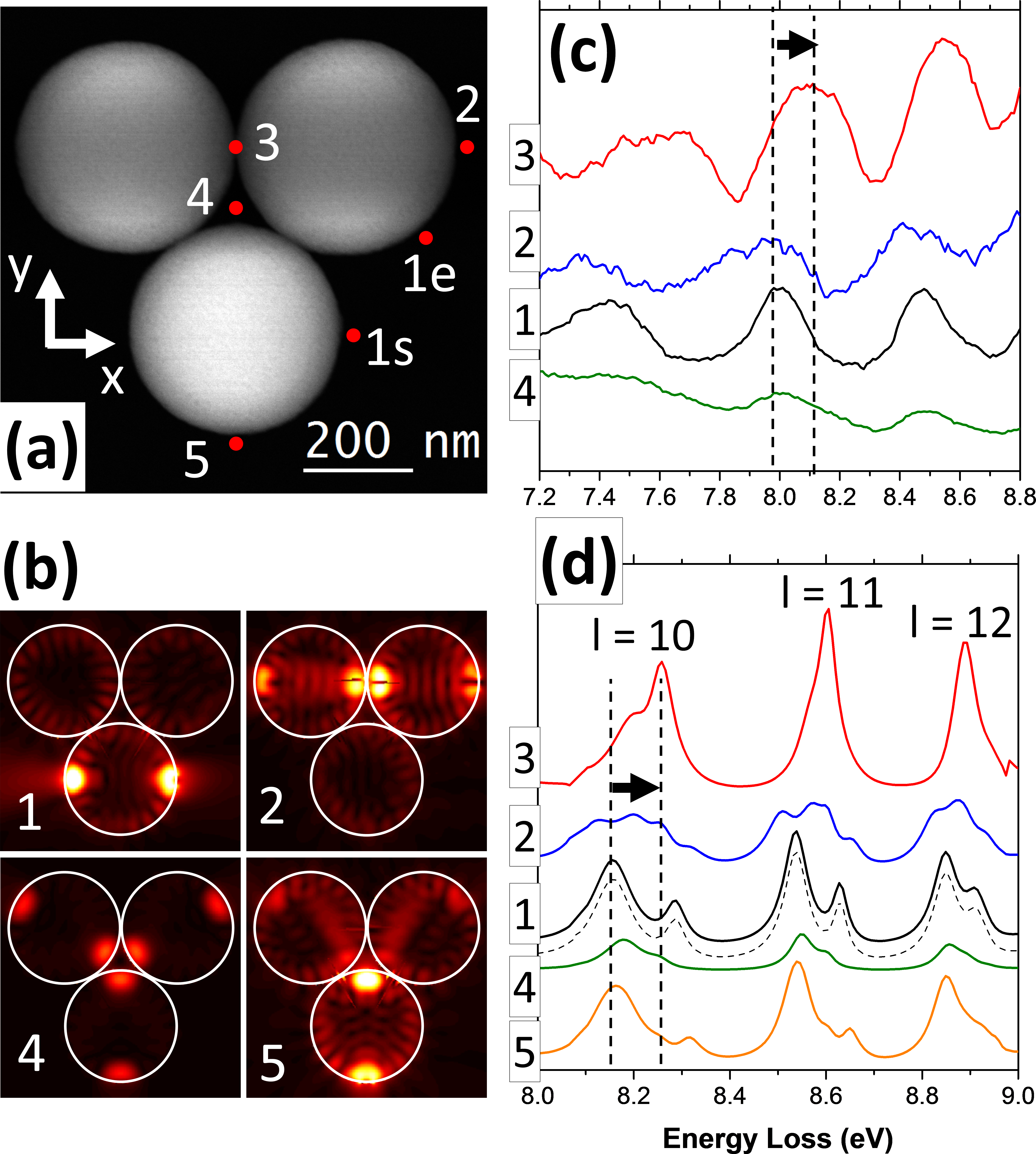}%
 \caption{\label{} (a) STEM image of a silica-sphere trimer with electron probe locations for $\vert H \vert$ maps displayed in (b), experimental spectra displayed in (c) and simulated spectra displayed in (d).  A simulated spectrum for an individual sphere is shown as a dotted line in (d).  $\vert H \vert$ maps are displayed on the \textit{x-y} plane that includes the centers of the spheres.  Locations 1, 2 and 5 are $\sim$5nm from the sphere's surface}
 \end{figure}

Figure 3 presents both experimental and simulated results for a silica-sphere trimer, as shown on the image in (a) along with probe locations for each energy-loss spectrum.  Location 1s is for the simulated spectrum, and basically corresponds geometrically to the experimental location 1e.   For location 1, both experimental (c) and simulated (d) spectra resemble the one for an isolated sphere.  As was the case with a dimer, peaks in spectra broadens for location 2 and are blue-shifted for location 3.  The energy shift for a peak at $\sim$8eV is $\sim$0.10 eV in both the experiments and the simulations (see arrows in (c) and (d)).  Finally, location 4, at the center of the trimer, displays peak maxima close to the ones for location 1, but with much reduced intensity due to the distance between the electron probe and the spheres’ surfaces.  

Figure 3(b) displays $\vert H \vert$ distributions on an \textit{x-y} plane that includes the centers of the three spheres.  For location 1, fields are confined mostly to one sphere.  For location 2 (and also 3, not shown), the fields are contained mainly within two spheres, but also extends slightly into the third sphere of the trimer, producing spectra that differs (slightly) from the ones for a dimer.  If the electron probe is located at the center of the trimer, location 4, fields are expected in the three spheres.  This is not necessary an indication of coupling.  But if we position the probe at location 5, it is clear that fields are observed inside the three spheres, and suggest that coupling will be expected for position 4 and 5, which explains the difference between their corresponding spectra (d) and the one for an isolated sphere. 
 
Coupling lifts the $(2l+1)$ degeneracy of optical modes in a sphere \cite{miyazaki2000}.  Experimentally, the multiresonance responses is often observed as envelopes of unresolved resonances.  In fluorescence optical spectroscopy \cite{mukaiyama1999}, TM optical resonances in a bisphere were found to split in two, with a barely detectable peak at higher wavelength and a dominant peak at lower wavelengths when compared to an isolated sphere.  These observations are consistent with the blue shift observed for probe location 3 for both dimers (Fig. 2) and trimers (Fig. 3).  With electron-beam spectroscopy, however, by positioning the electron probe at the nanoscale (e.g. location 2 for dimers and trimers), it is possible to access different sets of resonances. Furthermore, for our low-$l$ system ($l < 12$), coupling will include modes with $N \neq 1$ and $m \neq l$, and the fields will not be as concentrated close to the spheres’ equator.  This will complicate the interpretation of WGM coupling, and will increase the potential for coupling between multiple spheres that are not in a linear arrangement, as observed here for triangular trimers.  

In summary, we have demonstrated that electron-beam spectroscopy offers a path to probe optical response of (non-plasmonic) photonic molecules at the nanometer scale.  Whispering gallery modes in silica nanoparticles ($\sim$400 nm in diameter) were excited by 300keV electron probe, producing series of peaks in the ultraviolet domain of electron energy-loss spectra.  For silica-sphere dimers and trimers, we have observed experimentally that coupled modes are excited by incident electrons, and that different coupled modes were excited for different probe positions.  Even with limitations such as electron absorption in thick regions or possible beam damage, electron-beam spectroscopies offer unique access to optical responses with a spatial resolution that is much finer than the variation of the modes themselves.  

\textbf{Data availability}: The data that support the findings of this study are available from the corresponding author upon reasonable request.

\bibliography{mybib}

\end{document}